\documentclass[a4paper,twoside,twocolumn,english,showpacs]{revtex4}
\usepackage{times}
\usepackage[OT1]{fontenc}
\usepackage[latin1]{inputenc}
\usepackage{amsmath}
\usepackage{graphicx}
\usepackage{amssymb}

\makeatletter

\usepackage{babel}
\makeatother
\begin{document}

\title{Two point correlations of a trapped interacting Bose gas at finite temperature}

\author{A. Bezett}
\author{E. Toth}
\author{P. B. Blakie}

\affiliation{Jack Dodd Centre for Photonics and Ultra-Cold Atoms, Department of Physics, University of Otago, Dunedin, New Zealand}

\date{\today}


\begin{abstract}
We develop a computationally tractable method for calculating correlation functions of the finite temperature trapped Bose gas that includes the effects of s-wave interactions. Our approach uses a classical field method to model the low energy modes and treats the high energy modes using a Hartree-Fock description.
We present results of first and second order correlation functions, in position and momentum space, for an experimentally realistic system in the temperature range of $0.6T_c$ to $1.0T_c$. We also characterize the spatial coherence length of the system. Our theory should be applicable in the critical region where experiments are now able to measure first and second order correlations.

\end{abstract}
\maketitle

\section{Introduction}

Recent experimental developments in ultra-cold gases have seen an increase in interest in atomic correlation measurements analogous to the photon correlations observed in the landmark experiments of Hanbury-Brown and Twiss \cite{hbt}. In those experiments the second order correlation function (intensity correlations) of light was measured, revealing that photons from a thermal light source are bunched. Correlation measurements in atomic systems are of significant interest due to interaction effects and that both Bose and Fermi particles can be investigated. 
 
The first experiments with atoms by Yasuda \emph{et al.} \cite{yasuda} used a neutral (bosonic) atomic beam and confirmed atom bunching. Using ultra-cold Bose gases local high order correlations have  been inferred from 3-body decay rates \cite{threebody1,threebody2}, and  first order coherence has been studied using matter wave interference \cite{matterwave2,matterwave1} and Bragg spectroscopy \cite{AspectBragg, Ketterle}.

More recently there has been spectacular experimental progress in the spatially resolved measurement of second order correlations in both  bosonic and fermionic ultra-cold  gases \cite{aspect,esslinger,jeltes,folling,jin,Rom2006}. 
Two general approaches are used to make these measurements. One approach involves directly counting atoms \cite{aspect,esslinger,jeltes}, the other uses absorption imaging to measure density fluctuations \cite{folling,jin,Rom2006}. The applications of these measurements have included Bose and Fermi gases in harmonic traps \cite{aspect,jeltes,esslinger,jin}, and in optical lattices \cite{folling,Rom2006}.

Theoretical work has included the extension of optical definitions of coherence to atomic samples \cite{glauber,measure}, and examining how correlation measurements made after the sample has expanded from its confinement potential relate to those of the original \emph{in situ} system \cite{altman,TheoryAspect}. Theoretical predictions for correlations functions of the trapped Bose gas have included studies of the ideal case \cite{barnett,glauber,TheoryAspect} and interacting gases within mean-field descriptions \cite{glauber}.  
To date, the most comprehensive calculations of second order correlations in the interacting system used the Hartree Fock Bogoliubov formalism \cite{dodd}, however in that work only local correlations were investigated. A non-local extension of that formalism was used to calculate the first order correlation function of the quasi-2D Bose gas \cite{gies} \footnote{Our primary interest here is in the 3D Bose gas so we do not consider the rather extensive literature on quasi-1D and 2D gases (e.g. see \cite{Posazhennikova}).}.

Generally speaking, it is a considerable challenge to calculate these correlation functions in a manner that consistently includes the effect of interactions and harmonic confinement. Near the critical point, where a mean-field analysis is no longer valid, there are currently no reliable calculations that have been applied to realistic experimental systems. The primary purpose of this paper is to develop a classical field formalism for calculating non-local correlation functions that can describe the critical region. The necessity of such a theory is  motivated by two recent experiments: The  measurement of the critical exponent for the correlation length by the ETH Zurich group \cite{critical}; and the  measurement of the second order correlation function near the critical point by the Orsay group \cite{aspect}.   

Our formalism, developed in Sec. \ref{sec:Formalism}, is based upon a separation of the modes of the system according to their mean occupation: we use the Projected Gross Pitaevskii Equation (PGPE) to treat the  low energy, highly occupied modes, and a mean field Hartree Fock approach for the high energy, sparsely occupied modes. The PGPE approach non-perturbatively includes interactions between low energy modes of the gas and is applicable to the critical region. Indeed, its predictions for the shifts in critical temperature \cite{Davis06} are in good agreement with experimental measurements \cite{Gerbier04}. For the higher energy modes of the system, fluctuation effects are less important and a mean-field (Hartree Fock) approach is sufficient.

In Sec. \ref{sec:results} we present results of first and second order correlation functions for experimentally realistic systems, in position and momentum space coordinates.   The momentum space correlation function approximately corresponds to that measured after expansion (e.g. see \cite{TheoryAspect}). 
The approximate range of temperatures we investigate is $0.6T_c$ to $1.0T_c$, where $T_c$ is the critical temperature.   We also consider a coherence length observable and use this to compare the results of our method  to those of a (pure) mean-field approach.

\section{Formalism \label{sec:Formalism}}

 \subsection{System and correlation functions} 
We take our system to be described by  the second quantized Hamiltonian
\begin{eqnarray}
\hat{H} &=& \int  d\mathbf{r}\, \left\{\hat\Psi^{\dagger}(\mathbf{r}) \left(-\frac{\hbar^2}{2m}\nabla^2 + V_{trap}(\mathbf{r})\right)\hat\Psi(\mathbf{r})\right. \nonumber\\
&& \left.+ \frac{1}{2}U_0\hat\Psi^{\dagger}(\mathbf{r})\hat\Psi^{\dagger}(\mathbf{r})\hat\Psi(\mathbf{r})\hat\Psi(\mathbf{r})\right\} ,
\end{eqnarray}
where $\hat\Psi(\mathbf{r})$ is the quantum Bose field operator, and $U_0 = 4\pi\hbar^2a/m$ is the interaction strength, with $a$ the s-wave scattering length. The harmonic trapping potential is given by
\begin{equation}
V_{trap}(\mathbf{r}) = \frac{1}{2}m\left(\omega_x^2x^2+\omega_y^2y^2+\omega_z^2z^2\right),
\end{equation}
where $\{\omega_x,\omega_y,\omega_z\}$ are the trap frequencies.

\subsubsection{Position space correlation functions}
The density and unnormalised first and second order correlation functions (e.g. see \cite{glauber}) are defined as
\begin{eqnarray}
n(\mathbf{r}) &=& \langle\hat\Psi^{\dagger}(\mathbf{r})\hat\Psi(\mathbf{r})\rangle,\label{nfull}\\
G^1(\mathbf{r},\mathbf{r'}) &=& \langle\hat\Psi^{\dagger}(\mathbf{r'})\hat\Psi(\mathbf{r})\rangle, \label{G1full}\\
G^2(\mathbf{r},\mathbf{r'}) &=& \langle\hat\Psi^{\dagger}(\mathbf{r'})\hat\Psi^{\dagger}(\mathbf{r})\hat\Psi(\mathbf{r})\hat\Psi(\mathbf{r'})\rangle,  \label{G2full}
\end{eqnarray}
 where the averages are to be interpreted as thermal averages. We note that local first order correlations are equal to the density, i.e. $n(\mathbf{r})=G^1(\mathbf{r},\mathbf{r})$. 
For the purposes of interpreting particle correlations it is convenient to introduce the normalised versions of the correlation functions
\begin{eqnarray}
g^1(\mathbf{r},\mathbf{r'}) &=& \frac{G^1(\mathbf{r},\mathbf{r'})}{\sqrt{n(\mathbf{r'})n(\mathbf{r})}},\label{g1}\\
g^2(\mathbf{r},\mathbf{r'}) &=& \frac{G^2(\mathbf{r},\mathbf{r'})}{n(\mathbf{r'})n(\mathbf{r})}.
\end{eqnarray}

The first order correlation function describes phase fluctuations in the system, and as $G^1(\mathbf{r},\mathbf{r'})$ is the one-body density matrix of the system, off-diagonal long range order in this quantity is the defining characteristic of Bose-Einstein condensation \cite{Penrose1956}. 
The second order function is a measure of pair correlations in the system, for instance the atoms tendency to cluster (bunch) or separate (anti-bunch). Recent experiments in a superfluid Fermi gas have revealed non-local pairing through measurements of second order correlations \cite{jin}.

\subsubsection{Momentum space correlation functions}
{  Often the correlation functions for ultra-cold atom systems are measured after \emph{time of flight} expansion for time $t_{\rm exp}$. In this situation the measured correlations are proportional to the \emph{in situ} momentum correlations, with the relationship between final (observed) position $\mathbf{R}$ and \emph{in situ} momentum  given by $\mathbf{R}/t_{\rm exp}=\mathbf{p}/m$. For this reason we also develop our formalism to calculate momentum space correlations, defined in terms of the momentum field operator $\hat\Phi(\mathbf{p})$,  as
\begin{eqnarray}
n(\mathbf{p}) &=& \langle\hat\Phi^{\dagger}(\mathbf{p})\hat\Phi(\mathbf{p})\rangle,\label{pnfull}\\
G^1(\mathbf{p},\mathbf{p}') &=& \langle\hat\Phi^{\dagger}(\mathbf{p}')\hat\Phi(\mathbf{p})\rangle, \label{pG1full}\\
G^2(\mathbf{p},\mathbf{p}') &=& \langle\hat\Phi^{\dagger}(\mathbf{p}')\hat\Phi^{\dagger}(\mathbf{p})\hat\Phi(\mathbf{p})\hat\Phi(\mathbf{p}')\rangle.  \label{pG2full}
\end{eqnarray}
Many aspects of the formalism we present for position and momentum space are similar, and in what follows we focus primarily on giving a detailed derivation for the position case,  and will only comment on any important differences that arise in the momentum case. 

\subsection{Finite temperature formalism}
To describe the trapped Bose gas we split the full field operator into two parts, a classical field ($\psi_c$) representing the modes which are highly occupied (referred to as the coherent region), and a quantum field ($\hat\psi_i$) representing the sparsely populated modes (the incoherent region). 
The cutoff between the two will be taken to be where the occupation in the mode is around five particles.
\begin{equation}
\hat\Psi = \psi_c + \hat\psi_i\label{EqfieldOp}
\end{equation}
 We describe the details of this decomposition further in the following subsections.
There has been a significant body of work on applications of the classical field methods to  zero and finite temperature properties of Bose gases \cite{Davis2002a,DavisTemp,Davis2005a,Davis06,Simula2006a,Gardiner2002a,Gardiner2003a,Goral2001a,Davis2001b,Davis2001a,Lobo2004a,Marshall1999a,Norrie2004a,
Polkovnikov2004a,Sinatra2001a,Sinatra2002a,Steel1998a,Bradley,Blakie2004}. These studies  show that this splitting of the field operator can be done in a consistent manner and provides an accurate description of experiment (in particular see Ref. \cite{Davis06}).

Substituting expression (\ref{EqfieldOp})   into the  Eqs. (\ref{nfull})-(\ref{G2full})  gives
\begin{eqnarray}
n(\mathbf{r})&=&n_c(\mathbf{r}) + n_i(\mathbf{r}),\label{nfull2}\\
G^1(\mathbf{r},\mathbf{r'}) &=& G^1_c(\mathbf{r},\mathbf{r'}) + G^1_i(\mathbf{r},\mathbf{r'}),\label{G1full2}\\
G^2(\mathbf{r},\mathbf{r'}) &=& G^2_c(\mathbf{r},\mathbf{r'}) + G^2_i(\mathbf{r},\mathbf{r'}) + 2G^1_i (\mathbf{r},\mathbf{r'})G^1_c(\mathbf{r},\mathbf{r'})\nonumber\\
&& + n_i(\mathbf{r}) n_c(\mathbf{r'}) + n_i(\mathbf{r'}) n_c(\mathbf{r}),\label{G2full2}
\end{eqnarray}
where
\begin{eqnarray}
n_j(\mathbf{r}) &=& \langle\psi_j^{\dagger}(\mathbf{r})\psi_j(\mathbf{r})\rangle \\
G^1_j (\mathbf{r},\mathbf{r'})&=& \langle\psi_j^{\dagger}(\mathbf{r'})\psi_j(\mathbf{r})\rangle, \\
G^2_j (\mathbf{r},\mathbf{r'})&=& \langle\psi_j^{\dagger}(\mathbf{r'})\psi_j^{\dagger}(\mathbf{r})\psi_j(\mathbf{r})\psi_j(\mathbf{r'})\rangle, 
\end{eqnarray}
with $j=\{i,c\}$ for the incoherent and coherent regions respectively.

Several approximations have been made in deriving Eqs. (\ref{nfull2})-(\ref{G2full2}).
We assume that:
\begin{enumerate}
\item[(i)] the coherent and incoherent regions are uncorrelated, so that expectations of mixed terms of coherent and incoherent operators can be factorized, e.g. $\langle \hat\psi_i^\dagger\psi_c^*\psi_c\hat\psi_i\rangle=\langle \hat\psi_i^\dagger\hat\psi_i\rangle\langle\psi_c^*\psi_c\rangle$. For this to be a good approximation we require that our fields are expanded in a basis that approximately diagonalizes the problem at the energy cut off. For our purposes the single particle basis of the harmonic oscillator potential is satisfactory (also see \cite{Gardiner2003a}).
\item[(ii)] we can neglect the averages of single fields. We note that in the coherent region this is justified because we do not make the symmetry breaking approximation to describe the condensate.
\item[(iii)] we can neglect the averages of anomalous fields, e.g. $\langle  \psi_c\psi_c\rangle$. Within the classical field approximation the anomalous average of the coherent operator is zero, and the anomalous expectation of the incoherent field is also zero within the Hartree Fock approximation we use here.
\end{enumerate}

\subsection{Coherent Region}
Atoms in  the coherent region are treated with the Projected Gross Pitaevskii Equation  formalism \cite{Blakie2004}. 
In this approach the coherent field is expanded in a harmonic oscillator basis 
\begin{equation}
\psi_c(\mathbf{r},t)=\sum_{n\in  C}c_n(t)\varphi_n(\mathbf{r}),
\end{equation}
where $n$ represents the quantum numbers for the harmonic oscillator basis states $\{\varphi_n(\mathbf{r})\}$, and $c_n$ are time-dependent complex amplitudes. We consistently define the coherent region ($C$) by restricting the summation to only include oscillator states of (single particle) energy less than a predetermined energy cut off ($\epsilon_{\rm{cut}}$).
The equation of motion for the coherent field is the PGPE
\begin{eqnarray}
i\hbar\frac{\partial \psi_c ( \mathbf{{r}},t)}{\partial t} &=& \left(-\frac{\hbar^2}{2m}\nabla^2 + V_{trap}(\mathbf{{r}})\right)\psi_c(\mathbf{{r}},t) \nonumber\\
&&+ \mathcal{P}\left\{ U_0 |\psi_c(\mathbf{{r}},t)|^2\psi_c(\mathbf{{r}},t)\right\}, \label{PGPE}
\end{eqnarray}
where the projection operator ($\mathcal{P}$) formalises our basis set restriction to the coherent region.  The primary approximation used to arrive at the PGPE is to neglect interaction with the incoherent region (see \cite{Davis2001b}), which is expected to be well justified in the equilibrium case. 

Equation (\ref{PGPE}) can be efficiently and accurately simulated in the basis set representation using an appropriately chosen Gauss-Hermite quadrature (for more details see \cite{quad}).

The essence of the PGPE formalism is that the field evolution according to  Eq.~(\ref{PGPE}) is ergodic and equilibrium properties can be found by time averaging solutions. 
By ignoring the coupling to the incoherent modes, the coherent region forms a microcanonical system whose equilibrium states are specified by the constants of motion,  most  importantly energy 
 \begin{equation}
E_c=\int d^3\mathbf{r}\,\psi_c^*\left(-\frac{\hbar^2}{2m}\nabla^2 + V_{trap}(\mathbf{{r}}) +  \frac{U_0}{2} |\psi_c|^2\right)\psi_c,\end{equation}
  and the number of particles 
 \begin{equation}
 N_c = \int d^3\mathbf{r}\,|\psi_c(\mathbf{r})|^2.
 \end{equation}

Our simulation procedure is as follows: Initial states are prepared with the desired values of $E_c$ and $N_c$ } and evolved until the sample has thermalized. This is judged by monitoring when the time averages of parameters such as condensate fraction and temperature have settled down and fluctuate about steady state values \footnote{The procedure for determining condensate fraction, temperature and chemical potential is discussed in Ref. \cite{Blakie2004,DavisTemp,Davis2005a}.}. The values of $\mu$, $T$ and $n_c(\mathbf{r})$ which are extracted from the simulations (using methods discussed in \cite{DavisTemp}) are used further in calculations for the incoherent region.

Taking $N_s$ samples of the coherent field at  times $\{t_j\}$ after it has thermalized, we  evoke the ergodic hypothesis to evaluate the correlation functions as
\begin{eqnarray}
n_c(\mathbf{r}) &=& \frac{1}{N_s}\sum^{N_s}_{j=1}\left|\psi_c(\mathbf{r},t_j)\right|^2,\label{nc}\\
 G^1_c(\mathbf{r},\mathbf{r'}) &=& \frac{1}{N_s}\sum^{N_s}_{j=1}\psi^{*}_c(\mathbf{r},t_j)\psi_c(\mathbf{r'},t_j),\label{G1c}\\
 G^2_c(\mathbf{r},\mathbf{r'}) &=& \frac{1}{N_s}\sum^{N_s}_{j=1}\left|\psi_c(\mathbf{r'},t_j)\right|^2\left|\psi_c(\mathbf{r},t_j)\right|^2.\label{G2c}
\end{eqnarray}
Correlation functions in momentum space can be immediately evaluated using the procedure in (\ref{nc})-(\ref{G2c}), but  using the momentum field as given by 
\begin{equation}
\phi_c(\mathbf{p},t) = \frac{1}{(2\pi)^{3/2}} \int d^3 \mathbf{r}\, e^{-i\mathbf{p}\cdot\mathbf{r}/\hbar}\psi_c(\mathbf{r},t).
\end{equation}
 
\subsection{Incoherent Region}\label{incohregion}
Atoms of the incoherent region, i.e. those in high-energy sparsely occupied states, are treated with a semiclassical phase space approach based on the Hartree Fock meanfield analysis. For consistency we only apply this description to the volume of phase space complimentary to the coherent region.

The Wigner distribution for this region is given by
\begin{equation}
W_i(\mathbf{{r}},\mathbf{{p}}) = \frac{1}{h^3}\frac{1}{e^{{\beta(E(\mathbf{r},\mathbf{p}) - \mu)}} -1}, \label{density_i}
\end{equation}
where $\mu$ is the chemical potential and $\beta=1/k_BT$. 
The energy states are given by the Hartree-Fock expression
\begin{equation}
E(\mathbf{{r}},\mathbf{{p}}) = \frac{\mathbf{p}^2}{2m}+V_{trap}(\mathbf{{r}})   + 2U_0(n_{c}(\mathbf{{r}}) + n_{i}(\mathbf{{r}})),
\end{equation}
and the incoherent density is evaluated as
\begin{equation}
n_i(\mathbf{r})=\int^{\prime} d^3\mathbf{p}\,W_i(\mathbf{{r}},\mathbf{{p}}).\label{ni}
\end{equation}
The prime on the integral indicates a restriction in the region of  phase space coordinates we integrate over, i.e.
\begin{equation}
\frac{\mathbf{p}^2}{2m}+V_{trap}(\mathbf{{r}}) > \epsilon_{\rm{cut}}.
\end{equation}
This condition defines the incoherent region and ensures we do not double count contributions already in the coherent region.

The incoherent region formalism is important in determining the equilibrium properties of the complete system. It is only after self-consistently  determining the incoherent density using (\ref{density_i})-(\ref{ni}) that we are able to evaluate the total number of atoms in the incoherent region as $N_i=\int d^3\mathbf{r} \,n_i(\mathbf{r})$, and hence the total number of atoms for the system $N=N_c+N_i$.

The Wigner function is related to the first order correlation function by  \cite{glauber} 
\begin{equation}
G^1_i (\mathbf{r},\mathbf{r^{\prime}}) = \int^{\prime} d^3\mathbf{p}\,e^{-i\mathbf{p}\cdot(\mathbf{r} - \mathbf{r^{\prime}})/\hbar} W_i\left(\frac{\mathbf{r}+\mathbf{r}^{\prime}}{2},\mathbf{p}\right).
\end{equation}
 Since the phase space coordinates are a diagonal basis for our Hartree-Fock treatment of the incoherent region, we can obtain the second order correlation function as
\begin{equation}
G^2_i(\mathbf{r},\mathbf{r}^{\prime}) = G^1_i(\mathbf{r},\mathbf{r})G^1_i(\mathbf{r}^{\prime},\mathbf{r}^{\prime}) + |G^1_i(\mathbf{r},\mathbf{r}^{\prime})|^2.\label{scG2fromG1}
\end{equation}

The momentum space correlation functions for the incoherent region can be obtained in a similar manner to the position space case using the result
\begin{equation}
G^1_i (\mathbf{p},\mathbf{p^{\prime}}) = \int^{\prime} d^3\mathbf{r}\,e^{-i\mathbf{r}.(\mathbf{p} - \mathbf{p^{\prime}})/\hbar} W_i \left(\mathbf{r},\frac{\mathbf{p}+\mathbf{p^{\prime}}}{2}\right).
\end{equation}
Equation (\ref{scG2fromG1}) can also be applied to the momentum space result, and we do not repeat it here.

\section{Results}\label{sec:results}

In this section we present results of the application of our formalism to an ultra-cold Bose cloud. The system we consider consists of approximately $3\times10^5$ $\,^{87}$Rb atoms confined in an anisotropic harmonic trap of frequencies $\{\omega_x,\omega_y,\omega_z\}=2\pi\,\{1,1,\sqrt{8}\}\times40\,s^{-1}$.
We explore a temperature range of approximately $0.6T_c$  -- $1.0T_c$ to investigate the interplay of the thermal and condensate clouds, and to see how this affects the correlation functions. The analytic estimate of the critical temperature including finite size and mean-field shifts \cite{Giorgini1996a} gives $T_c\approx162$ nK for this system, however our results suggest the actual $T_c$ may be slightly lower, with   the system attaining  $\sim1\%$ condensate fraction at about 159 nK. 

 \begin{figure}
\includegraphics[width=3.4in, keepaspectratio]{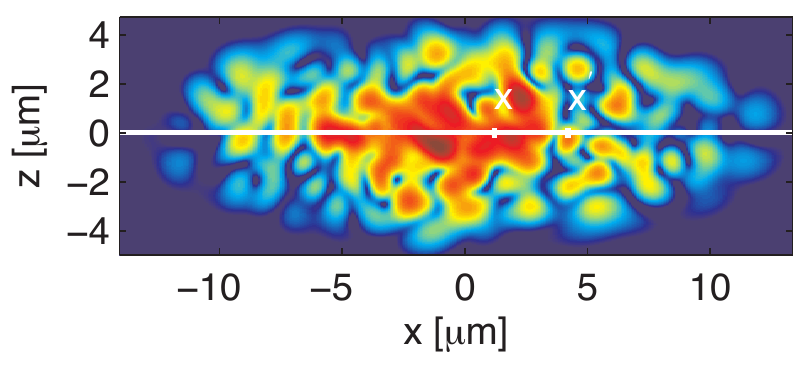}
 \caption{\label{correldiag} (color online) Instantaneous density slice of classical field in $xz$-plane showing the system and correlation measurement geometry. The $x$-axis is indicated (white line) and two points $x$ and $x^\prime$ are shown. The  field and density correlations between these points define $G_c^1(x,x^\prime)$ and $G_c^2(x,x^\prime)$ respectively. This result is for  a system of $3\times10^5$ $\,^{87}$Rb atoms in the trap described in the text at a temperature of $159$ nK. }
\end{figure}
For our case of a system with external confinement, the two-point correlation functions will depend on all coordinates \footnote{This is in contrast to the homogeneous case where translational invariance means that correlation functions only depend on the relative separation of the points e.g. $G^2_{\rm{hom}}(\mathbf{r},\mathbf{r}^\prime)=G^2_{\rm{hom}}(\mathbf{r}-\mathbf{r}^\prime)$.}.
Hence a complete characterization of these correlations in the 3D system requires six-dimensions.  The results we present here are for the correlation functions of the full 3D system for the case of both points lying on the $x$-axis, e.g. $G^1(\mathbf{r}\!=\!x\mathbf{\hat{x}},\mathbf{r'}\!=\!x'\mathbf{\hat{x}})$, where $\mathbf{\hat{x}}$ is the unit vector in $x$ direction, and this will now be abreviated to just $G^1(x,x')$. The geometry of this is indicated in Fig. \ref{correldiag}.
 
\subsection{Position space correlations}

\begin{widetext}

 \begin{figure}
\includegraphics[width=6in, keepaspectratio]{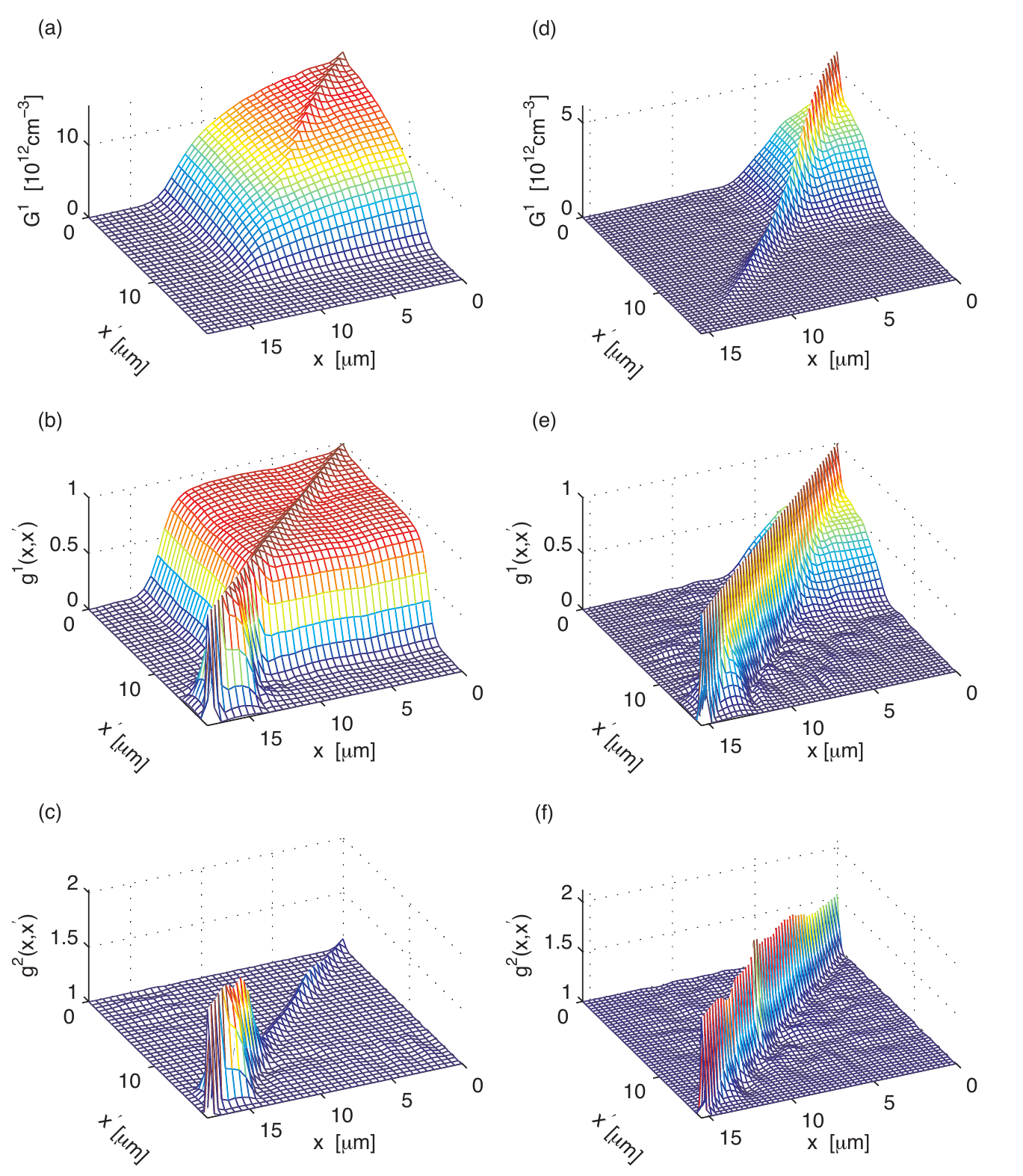}
 \caption{\label{g1162} (color online) Position space correlation functions  of a harmonically trapped Bose gas. (a)-(c) at $93nK$ and (d)-(f) $159nK$. Other parameters: (a)-(c) $N_{cond}=153\times10^3$ (condensate number), $N_c = 1.8\times10^5$, $N_i = 1.2\times10^5$, $\epsilon_{cut} = 44\hbar\omega_z$, ${E_c} = {N_c}\times21.8\hbar\omega_z$; (d)-(f) $N_{cond}=3540$, $N_c = 1.8\times10^4$, $N_i = 2.9\times10^5$, $\epsilon_{cut} = 36\hbar\omega_z$, ${E_c} = {N_c}\times20.5\hbar\omega_z$.  }
\end{figure}

\end{widetext}

Figure \ref{g1162} shows the one body density matrix ($G^1(x,x^\prime)$) and the normalized first and second order correlation functions for temperatures $93$ nK  [(a)-(c)] and $159$ nK  [(d)-(f)].

In the one body density matrix [Fig. \ref{g1162} (a) and (d)], two distinct features are clearly apparent. (i) A narrow ridge runs down the diagonal. The peak values of this ridge  gives the density of the system (recall $n(x)=G^1(x,x)$). We interpret the width of the ridge as the length scale over which phase coherence decays for the thermal component of the system. (ii) The broad background feature represents the \emph{off diagonal long range order} in the system, which is the defining characteristic of condensation in interacting systems \cite{Penrose1956}.  For the lower temperature result [Fig. \ref{g1162} (a)] the long range order dominates due to the large condensate fraction. In contrast, in the higher temperature 
case  [Fig. \ref{g1162} (d)] the thermal component is much more significant compared to the condensate, which is smaller in peak density and spatial extent. This result is close to the critical point and the thermal component has a substantial density even at the trap center.
 
We note that while the background feature arises entirely from the coherent region, the ridge has contributions from both coherent and incoherent regions.

In the normalised first order correlations [Fig. \ref{g1162} (b) and (e)], the background and ridge features are still apparent. 
However, normalization emphasizes the thermal component's contribution at large distances from the trap center. The ridge peak value is now unity, as is clear from Eq. (\ref{g1}).

The broadened feature seen in the ridge at large $x$ ($x\!\approx\!x^\prime\!\approx\!15\mu m$) is an artifact that arises from the limitations of our semiclassical description for the incoherent region. 
The position where this occurs corresponds to the classical turning point  for the energy cutoff used to define the coherent region \footnote{The classical turning point for the harmonic trapping is defined as $V_{trap}(\mathbf{r})=\epsilon_{\rm{cut}}$.}. Beyond this point all the modes of the coherent region are evanescent, and in examination of $G^1_c(x,x')$ this manifests as an apparent long range order. The local nature of the semiclassical approximation means that $G^1_i(x,x')$ does not cancel this feature, as we would expect in a more complete (wave) treatment of the incoherent region. 
With reference to Fig. \ref{g1162}(a) and (d) we note that this artifact occurs in a low density region and should have a minor  effect when results are averaged over the whole system.

We now consider the second order correlations shown in Fig. \ref{g1162} (c) and (f), which show a ridge, but no background feature. The height of the ridge varies from a value of slightly above one up to two.  
Comparing the second order results with the respective first order results, we see that in locations where the density is dominated by the condensate, the value of $g^2(x,x)$ is  suppressed  from the maximum value of two. 

These observations are consistent with the well known behavior of $g^2(x,x')$ for photons. For the case of an ideal laser $g^2(x,x')=1$, whereas for a thermal light source photon bunching  occurs with $g^2(x,x)=2$.

We note that the artifact seen in  Fig. \ref{g1162} (b) and (e), and discussed above, is also seen in these results.

\subsection{Results of correlation functions in momentum space}

\begin{widetext}

\begin{figure}
\includegraphics[width=6in, keepaspectratio]{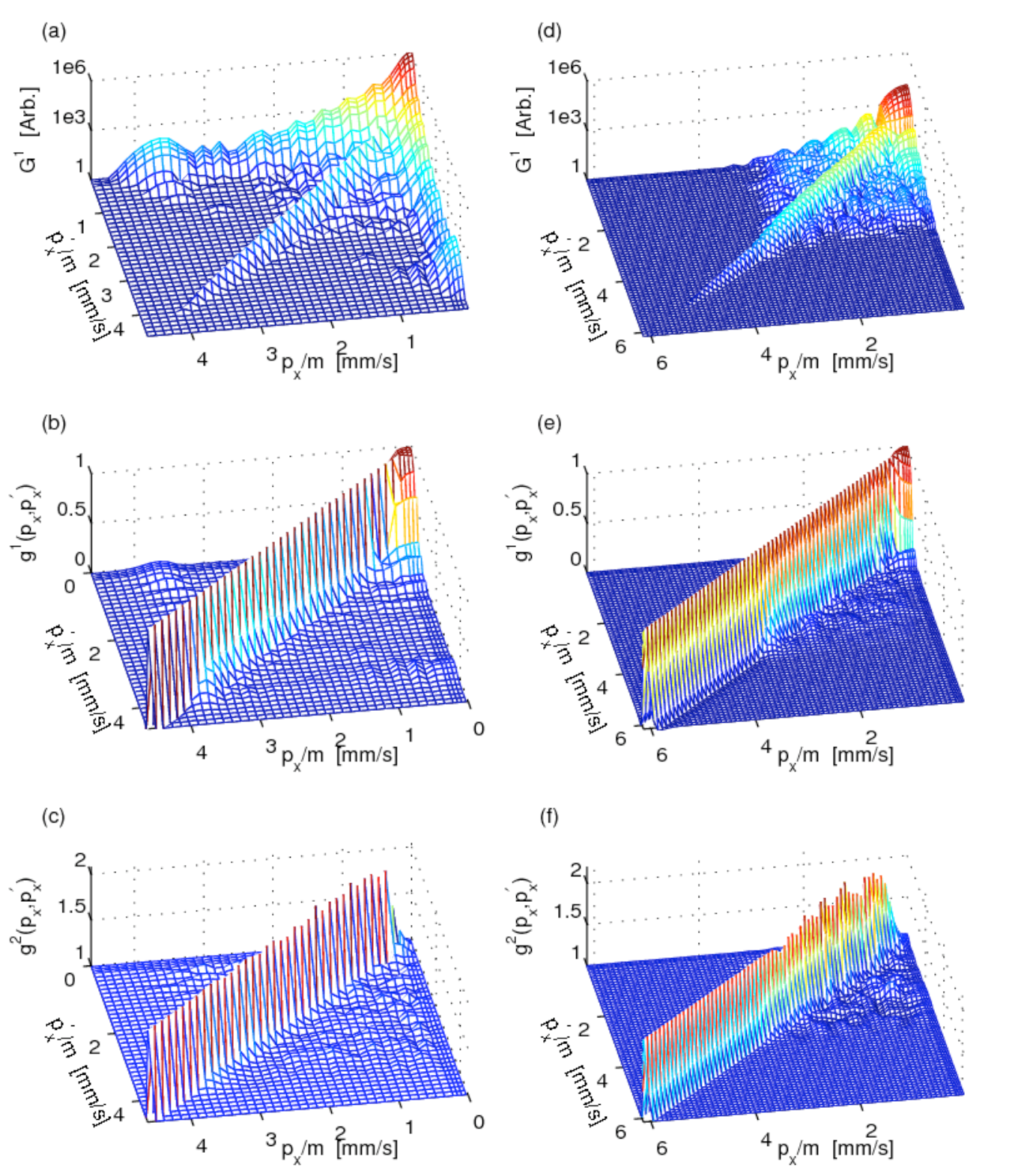}
 \caption{\label{mtmg1162} (color online) Momentum space correlation functions of a harmonically trapped Bose gas. (a)-(c) at $93nK$ and (d)-(f) $159nK$. Other parameters: (a)-(c) $N_{cond}=153\times10^3$, $N_c = 1.8\times10^5$, $N_i = 1.2\times10^5$, $\epsilon_{cut} = 44\hbar\omega_z$, ${E_c} = {N_c}\times21.8\hbar\omega_z$; (d)-(f) $N_{cond}=3540$, $N_c = 1.8\times10^4$, $N_i = 2.9\times10^5$, $\epsilon_{cut} = 36\hbar\omega_z$, ${E_c} = {N_c}\times20.5\hbar\omega_z$.  }
\end{figure}

\end{widetext}

Figure \ref{mtmg1162} shows the one body density matrix ($G^1(p_x,p_x^\prime)$) and the normalized first and second order correlation functions in momentum space for temperatures $93$ nK [(a)-(c)] and $159$ nK [(d)-(f)]. 
The dominant feature in 
$G^1(p_x,p_x^\prime)$ 
is a sharp spike in momentum which is a signature of condensation. 
 Indeed, we enhance the appearance of the thermal component in Fig. \ref{mtmg1162} (a) and (d) by using a logarithmic scale in which the density varies by six orders of magnitude.
In contrast, the position case [Figs. \ref{g1162}(a) and (d)] has a clearly discernible thermal component on a linear scale.

Comparing the low and high temperature results in Figures \ref{mtmg1162}(a) and (d) respectively, we observe the peak density of the condensate spike to vary by approximately two orders of magnitude, whereas for the same temperature change the position space density is observed to only change  by a factor of around two. This observation emphasizes that condensation is in some sense a momentum space phenomenon, but also marks a rather important difference between the position and momentum space correlation functions. The cross-like structure extending along the $p_x$ and $p_x^\prime$ axes in Fig. \ref{mtmg1162} (a) is due to the enhancement that the large condensate momentum space peak provides [this feature is negligible after normalization -- see Fig. \ref{mtmg1162} (b)].  

The normalized versions of the correlation functions more clearly reveals the non-condensed modes in the system.
 Due to the massive contrast between condensate and non-condensate modes in momentum space, experimental measurement of these correlations will likely prove challenging. Indeed, in Ref. \cite{aspect} results were restricted to above $T_c$ due to saturation issues with the detector when a condensate was present.

\subsection{Coherence length} 
The coherence length is a measure of first order correlations that specifies the typical length scale over which reproducible interference fringes might be expected. Recent experiments \cite{critical} have made detailed measurements of the coherence between atoms at different spatial locations within a trapped Bose gas using RF-fields to output couple the atoms.
This approach has the advantage that it avoids volume averaging which tends to smear features of the correlation functions (e.g. see \cite{glauber}).
This motivates us to consider an \emph{on-axis} coherence length defined as
\begin{equation}
{L}^2_x = \frac{\int dx\,dx'\, | G^1(x,x')|^2 (x-x')^2}{2 \int dx\,dx'\,| G^1(x,x')|^2}
\end{equation}
 which will be used to compare results at different temperatures. This expression is similar in form to the coherence length defined by  Barnett \emph{et al.} \cite{barnett} (also see Ref. \cite{measuresofCoh}), but without volume averaging over the whole system, and should be more appropriate for the aforementioned experiments.
 For reference, the uniform Boltzmann gas with $G^1(x,x')\sim\exp\left(-\pi |x\!-\!x'|^2/\lambda_{db}^2\right)$, gives ${L}_x=\lambda_{db}/\sqrt{8\pi}$ where $\lambda_{db}=h/\sqrt{2\pi mkT}$ is the thermal de Broglie wavelength.

\begin{figure}
\includegraphics[width=3.4in, keepaspectratio]{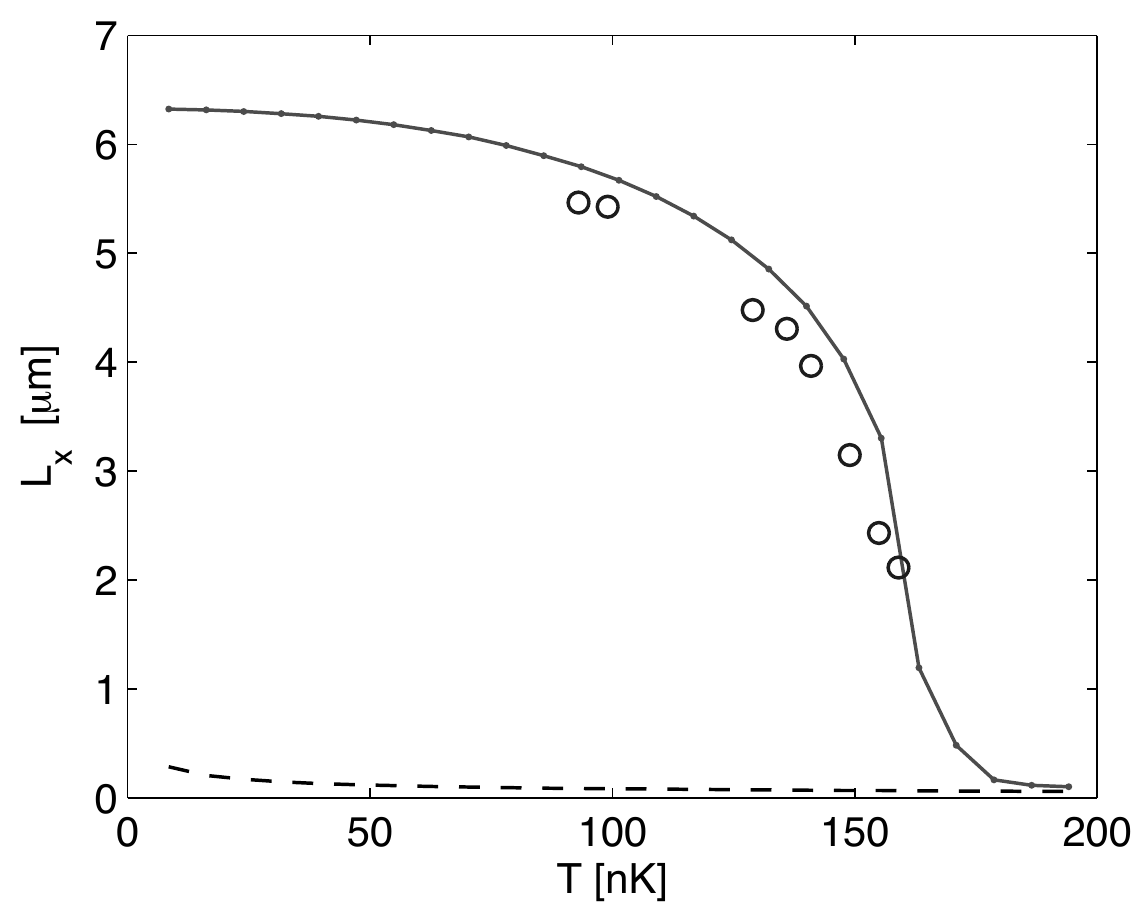}
\caption{\label{Lnonint} Coherence length ${L}_x$ calculated using classical field approach (circles), and semiclassical HFBP theory (line with dots). For reference $\lambda_{db}/\sqrt{8\pi}$ is shown as the dashed line. 
 System parameters as in Fig. \ref{g1162}}
\end{figure}

Figure \ref{Lnonint} shows  the behaviour of the on-axis coherence length for the same system considered in the previous section.
We note that this expression for the coherence length is dominated by the condensate mode, such that when the condensate fraction is appreciable using the full $G^1(x,x^\prime)$ or the condensate contribution $G_{\rm{cond}}^1(x,x^\prime)\equiv\Psi_0^*(x)\Psi_0(x^\prime)$, where $\Psi_0(x)$ is the condensate mode, yields almost identical results. Our results show the coherence length decreases steadily as the critical temperature is approached from below. For comparison we have also calculated $L_x$ using a self-consistent meanfield approach. Briefly, this approach describes the condensate (for $T<T_c$) using the Thomas-Fermi approximation and the non-condensate using Hartree-Fock Bogoliubov Popov (HFBP) theory in the semiclassical limit (e.g. see Ref. \cite{Giorgini1996a}). The first order coherence function is then constructed as $G_{\rm{HFBP}}^1(x,x^\prime)=G_{\rm{cond}}^1(x,x^\prime)+G_{\rm{th}}^1(x,x^\prime)$, where the thermal coherence function $G_{\rm{th}}^1(x,x^\prime)$ is obtained via the Wigner function as we have explained earlier for the incoherent region (see Sec. \ref{incohregion} but with $\epsilon_{\rm{cut}}\to0$, also see Ref. \cite{glauber}).

The results of Fig. \ref{Lnonint} indicate that the value of ${L}_x$ calculated using meanfield theory is always greater than the value obtained from the PGPE simulations for the temperatures considered. This primarily occurs because the equilibrium condensate fraction in the meanfield results is greater than the PGPE case at the same temperature\footnote{Discrepancies between the HFBP and PGPE calculations for the system equilibrium properties, in particular condensate fraction, may be due to limitations of the semiclassical approximation in addition to beyond meanfield effects.}.  Nevertheless, while there is reasonable agreement between these theories for $T<T_c$, as $T\to T_c$ their behaviour appears to be rather different, with the classical field prediction for $L_x$ decreasing much less rapidly on the approach to the critical point. We note that for the highest temperature  classical field result (i.e. $T=159$ nK) the condensate fraction is $\sim 1\%$ and the many other low-lying modes will necessarily play an important role in the coherence properties of the system in this regime. A complete investigation of this behavior is probably best done via the correlation length, $\xi$, defined such that $G^1(x,0)\propto x^{-1}\exp(-x/\xi)$, valid for $x\gtrsim \lambda_{db}$, and will be the subject of future work.

\section{Conclusions}
 
We have presented the theoretical development of an efficient, computationally tractable method for calculating correlation functions of the finite temperature trapped Bose gas in position and momentum space. Our method is robust and we have calculated results in a regime comparable to typical experiments.

Our results show the generic characteristics of these functions, emphasize the striking differences between their behavior in position and momentum space, and reveal the interplay between condensate and thermal components of the system. 
Most current experiments measuring two point correlations use expansion imaging, which yields information on the momentum space correlations. We have shown that when a condensate is present, this sort of measurement will require sensitivity to many orders of magnitude variation in atom density to yield accurate information on the condensate and thermal components. On the other hand, \emph{in situ} measurements, e.g. by output coupling \cite{critical}, may allow direct access to position space correlation measurements and require far less sensitivity to density variation.

Having developed a general formalism in this paper, we are currently working on applications to several systems of current interest. The primary strength of our method is that the low-energy classical field treatment is valid in regimes with strong (classical) fluctuations. We are investigating the application of our theory to such fluctuating regimes, motivated by two recent experiments:
(i) The change of first order coherence from exponential to algebraic decay has been used as a signature of the Berezinskii-Kosterlitz-Thouless transition in trapped quasi-2D Bose gas \cite{matterwave1,Polkovnikov}.
(ii) The measurement of the critical exponent in the trapped three-dimensional Bose Einstein condensation transition by Donner \emph {et al.} \cite{critical}, yielding a value of $\nu=0.67\pm0.13$ (compared to the ideal harmonically trapped case of $\nu=0.5$).

\section*{Acknowledgments}
A.B. and E.T. acknowledge support of  Top Achiever Doctoral Scholarships.
P.B.B. acknowledges support from the Marsden Fund of New Zealand, the University of Otago and the New Zealand Foundation for Research Science and Technology under the contract NERF-UOOX0703: Quantum Technologies, and useful discussions with MJ Davis.

\bibliographystyle{apsrev}
\bibliography{PGPE}

\end{document}